\documentclass[twocolumn,showpacs,preprintnumbers,amsmath,amssymb]{revtex4}
\usepackage{amsmath}
\usepackage{amssymb}
\usepackage{graphicx}
\usepackage{bm}
\usepackage{epstopdf}
\usepackage{slashed}
\usepackage{color}

\begin{document}

\title {Field sources in a CPT-even Lorentz-violation Maxwell electrodynamics}

\author{L. H. C. Borges}
\email{luizhenriqueunifei@yahoo.com.br}

\affiliation{UNESP - Campus de Guaratinguet\'a - DFQ, Avenida Dr. Ariberto Pereira da Cunha 333, CEP 12516-410, Guaratinguet\'a, SP, Brazil}

\author{F. A. Barone}
\email{fbarone@unifei.edu.br}

\affiliation{IFQ - Universidade Federal de Itajub\'a, Av. BPS 1303, Pinheirinho, Caixa Postal 50, 37500-903, Itajub\'a, MG, Brazil}

\begin{abstract}
This paper is dedicated to the study of interactions between external sources for the electromagnetic field in a model which exhibits Lorentz symmetry breaking. We investigate such interactions in the CPT-even photon sector of the Standard Model Extension (SME), where the Lorentz symmetry breaking is caused by a background tensor $K_{(F)\alpha\beta\sigma\tau}$. Since the background tensor is very tiny, we treat it perturbatively up to first order and we focus on physical phenomena which have no counterpart in Maxwell electrodynamics. We consider effects related to field sources describing point-like charges, straight line currents and Dirac strings. We also investigate the so called Aharonov-Bohm bound states in a Lorentz-symmetry breaking scenario. We use atomic experimental data to verify if we could impose upper bounds to the Lorentz-symmetry breaking parameters involved. We also use some overestimated constrains for
the Lorentz-symmetry breaking parameters in order to investigate if the obtained results could be  
relevant for condensed matter systems.
\end{abstract}

\maketitle
%
\section{Introduction}
\label{I}

The Lorentz-symmetry breaking is a promising pros\-pect for experimental detection of new physics at the Planck scale, and the field theoretic approach turns out to be the best way to describe the physics in this context \cite{LV1,LV2}. The comprehensive realistic effective field theory for Lorentz and CPT violation incorporating both the Standard Model (SM) and General Relativity is the Standard-Model Extension (SME) \cite{LV3,LV4}. The SME is a general framework which helps to develop investigations concerning the breaking of Lorentz and CPT symmetries in attainable energy scales.

The photon sector of the SME is composed by a CPT-even and a CPT-odd part. The CPT-odd sector is given by the Carroll-Field-Jackiw model $\sim k_{\mu}\varepsilon^{\mu\nu\rho\sigma}A_{\nu}F_{\rho\sigma}$, whose properties were investigated, firstly, in Ref.\cite{CFJ}. The CPT-even sector is given by the term $\sim F^{\alpha\beta}K_{(F)\alpha\beta\sigma\tau}F^{\sigma\tau}$, where the background tensor $K_{(F)\alpha\beta\sigma\tau}$ has nineteen independents components, with ten of them being sensitive to birefringence and nine being nonbirefringent \cite{CPT1,CPT2}. The nonbirefringent components can be studied  from experimental tests involving the Cherenkov radiation \cite{EXP1,EXP2,EXP3,EXP4}. The birefringent components are strongly limited through astrophysical tests involving data of cosmological sources \cite{EXP5}.

The photon sector of the SME has been extensively investigated in literature. We can mention, for instance, the vacuum emission of Cherenkov radiation \cite{CHR1,CHR2,CHR3}, modifications on the Casimir effect \cite{Casimir,Casimir2}, studies concerning the radiative generation of the CFJ term \cite{GP1,GP2,GP3,GP4,GP5}, modifications in the dispersion relations \cite{Wave,ClaE2,Wave2,CPTe1,CPTe2,CPTe6,CPTe7,CPTe8,CPTe12}, effects on the classical electrodynamics \cite{ClaE1,Fontes,CPTe9,CPTe10,CPTe11}, effects induced in the hydrogen atom \cite{HydrLV}, the photon field quantization \cite{GRP1,GRP2,GRP3}, effects of Aharonov-Bohm type \cite{AB1,AB2,AB3} and so on. 

One of the most fundamental questions one can make about models with Abelian gauge fields which exhibit explicit Lorentz-symmetry breaking concerns on the physical phenomena produced by the presence of external field sources, mainly on the phenomena with no counterpart in the Maxwell electrodynamics. Studies of this kind were, recently, performed in literature in reference \cite{Fontes}, where it was considered a Lorentz-symmetry breaking from the CPT-even sector of the SME, and in reference \cite{Fontes2}, where it was considered a Lorentz-symmetry breaking model with higher order derivatives in the field variables. Field sources in Lorentz-symmetry breaking scenarios were also considered for the graviton field \cite{LuizDenis}. 

In this paper we search for new effects produced by the presence of field sources in the CPT-even photon sector of the SME, where the background tensor $K_{(F)\alpha\beta\sigma\tau}$ is responsible for introducing the Lorentz-symmetry breaking. We perform our analysis in a similar manner that was employed in Ref's \cite{Fontes,Fontes2} and we treat the background tensor up to the leading order by using standard perturbative methods of Quantum Field Theory. We show that our results are in agreement with the ones of Ref. \cite{Fontes} if we write the background tensor $K_{(F)\alpha\beta\sigma\tau}$ as a function of a single background vector $v^{\mu}$ (as in Ref. \cite{Fontes}) and if we take the results of Ref. \cite{Fontes} in lowest order in $v^{\mu}$. We also show that new results, not yet explored in the literature, emerge from the considered model when we take examples where the background tensor cannot be written as a function of a single background vector.

Specifically, we show that it emerges a spontaneous torque on a classical electromagnetic dipole and an interaction between a steady straight line current and a point-like charge. We investigate some phenomena due to the presence of a Dirac string and show that the string can interact with a point charge as well as with a straight steady line current in the Lorentz-symmetry breaking scenario considered. We compute the electromagnetic field produced by the string.

In connection with the results related to Dirac strings, we make a study concerning the Aharonov-Bohm bound states of the $2$-dimensional quantum rigid rotor. We obtain the corrections to the energy levels, of this system, induced by the presence of the background tensor for an ilustrative and specific example.

Some of the results obtained along the paper are compared with experimental atomic data in order to stablish upper bounds to the Lorentz-symmetry breaking parameters. We
also make numerical estimates in order to investigate
the relevance of the obtained results in condensed matter
systems.

The paper is structured as follows; in  Section (\ref{II}) we describe some general aspects of the model used along the paper and compute the contribution, due to the sources, to the ground state energy of the system. In Section (\ref{III}) we consider effects due to the presence of point-like stationary charges. In Section (\ref{IV}) we obtain the interaction energy between a steady line current and a point-like stationary charge. Section (\ref{V}) is dedicated to the study of physical phenomena due to the presence of one Dirac string. In the section (\ref{Aharonov}) we calculate the corrections to the so called Aharonov-Bohm bound states of a $2$-dimensional quantum rigid rotor. Section (\ref{conclusoes}) is dedicated to our final remarks and conclusions. 

Along the paper we shall deal with models in $3+1$ dimensional space-time and use Minkowski coordinates with diagonal metric $\eta^{\mu\nu}=(+,-,-,-)$.

\section{The model}
\label{II}
The CPT-even part of the SME gauge sector is described by the following Lagrangian
\begin{eqnarray}
\label{modeloEM}
{\cal L}=-\frac{1}{4}F_{\mu\nu}F^{\mu\nu}-\frac{1}{2\xi}\left(\partial_{\mu}A^{\mu}\right)^{2}-\frac{1}{8}K_{(F)\alpha\beta\sigma\tau}F^{\sigma\tau}F^{\alpha\beta} \nonumber\\
+J^{\mu}A_{\mu} \ ,
\end{eqnarray}
where $A^{\mu}$ is the photon field, $F^{\mu\nu}=\partial^{\mu}A^{\nu}-\partial^{\nu}A^{\mu}$ is the field strength, $J^{\mu}$ is the external source, $\xi$ is a gauge parameter and $K_{(F)\alpha\beta\sigma\tau}$ is a dimensionless background tensor responsible for the Lorentz-symmetry breaking. The tensor $K_{(F)\alpha\beta\sigma\tau}$ has the same symmetries as the Riemann tensor
\begin{eqnarray}
\label{simetriasQ}
K_{(F)\alpha\beta\sigma\tau}=K_{(F)\sigma\tau\alpha\beta} \ ,\nonumber\\ 
K_{(F)\alpha\beta\sigma\tau}=-K_{(F)\beta\alpha\sigma\tau}=-K_{(F)\alpha\beta\tau\sigma}=K_{(F)\beta\alpha\tau\sigma}\ ,\nonumber\\ 
K_{(F)\alpha\beta\sigma\tau}+K_{(F)\alpha\tau\beta\sigma}+K_{(F)\alpha\sigma\tau\beta}=0\ ,
\end{eqnarray}
and a null double trace 
\begin{eqnarray}
\label{dtrace}
K^{\mu\nu}_{(F)\  \mu\nu}=0.
\end{eqnarray}

The generating functional for the model (\ref{modeloEM}) can be written in the following way \cite{Zee}
\begin{eqnarray}
\label{lae}
{\cal Z}&=&\int DA\exp\left(\frac{i}{2}\int d^{4}x \ A^{\mu}K_{(F)\mu\alpha\nu\beta}\partial^{\alpha}\partial^{\beta}A^{\nu}\right)\nonumber\\
&
&\times\exp\Biggl\{i\int d^{4}x\Biggl[\frac{1}{2}A_{\mu}\Biggl(\eta^{\mu\nu}\partial_{\alpha}\partial^{\alpha}\nonumber\\
&
&-\left(1-\frac{1}{\xi}\right)\partial^{\mu}\partial^{\nu}\Biggr)A_{\nu} +J^{\mu}A_{\mu}\Biggr]\Biggr\} \ .
\end{eqnarray}

Since the background tensor is very tiny, let us treat it perturbatively up to first order.

By using standard methods of quantum field theory, we perform the substitution \cite{Zee}
\begin{equation}
A^{\mu}\left(x\right)\rightarrow\frac{1}{i}\frac{\delta}{\delta J_{\mu}\left(x\right)} \ ,
\end{equation}
for the Lorentz violating term (first line in Eq. (\ref{lae})) and expand it, up to first order, in such a way that
\begin{eqnarray}
\label{lah}
{\cal Z}=\left[1-\frac{i}{2}\int d^{4}x\left(\frac{\delta}{\delta J^{\mu}}\right)K_{(F)}^{\mu\alpha\nu\beta}\partial_{\alpha}\partial_{\beta}\left(\frac{\delta}{\delta J^{\nu}}\right)\right]{\cal Z}_{0} ,
\end{eqnarray}
where $\delta/\delta J_{\mu}\left(x\right)$ stands for functional derivative with respect to the external source, ${\cal Z}_{0}$ is the free generating functional for the electromagnetic field \cite{Zee}
\begin{eqnarray}
\label{laj}
{\cal Z}_{0}=\exp\left[-\frac{i}{2}\int\int d^{4}y \ d^{4}z\ J^{\rho}\left(y\right)\Delta_{\rho\gamma}(y,z)J^{\gamma}\left(z\right)\right] \ ,
\end{eqnarray}
and $\Delta_{\rho\gamma}(y,z)$ is the free photon propagator 
\begin{eqnarray}
\label{lao}
\Delta_{\rho\gamma}(y,z)=-\int\frac{d^{4}p}{(2\pi)^{4}}\frac{1}{p^{2}}\left[\eta_{\rho\gamma}-(1-\xi)\frac{p_{\rho}p_{\gamma}}{p^{2}}\right]\nonumber\\
\times\exp[-ip\cdot(y-z)] \ .
\end{eqnarray}

The ground state energy of any quantum system can be written in terms of the generating functional, as follows \cite{Zee} 
\begin{eqnarray}
\label{lai}
E=\frac{i}{T}\ln{\cal Z} \ ,
\end{eqnarray}
where $T$ is the time variable and it is implicit the limit $T\rightarrow\infty$.

Substituting Eq. (\ref{laj}) in (\ref{lah}), acting with the functional derivatives and using Eq. (\ref{lai}) we obtain
\begin{eqnarray}
\label{IEES}
E&=&\frac{i}{T}\ln {\cal Z}_{0}-\frac{i}{2T} \ K_{(F)\mu\alpha\nu\beta}\int d^{4}x \ \partial^{\alpha}\partial^{\beta}\Delta^{\mu\nu}\left(x,x\right)\nonumber\\
&
&-\frac{1}{2T} \ K_{(F)\mu\alpha\nu\beta}\int\int d^{4}y \ d^{4}z \ J_{\rho}\left(y\right)J_{\gamma}\left(z\right)\nonumber\\
&
&\times\int d^{4}x \ \Bigl(\partial^{\alpha}\partial^{\beta}\Delta^{\nu\rho}\left(x,y\right)\Bigr)
\Delta^{\mu\gamma}\left(z,x\right) \ ,
\end{eqnarray}
where we used the approximation $\ln(1+x)\cong x$. 

The first term on right hand side of Eq. (\ref{IEES}) is the energy of the field in the absence of the background tensor, which is well known in literature \cite{BaroneHidalgo1,BaroneHidalgo2}. We can neglected the second term on the right hand side of Eq. (\ref{IEES}) since it does not depend on the external sources and, therefore, does not contribute to the interaction energy between field sources. The third term is a contribution to the interaction energy due to the Lorentz-symmetry breaking. Then, the energy of our quantum system due to the presence of the external source is given by the sum of the first and third terms in Eq. (\ref{IEES}).

Substituting Eq. (\ref{lao}) in the third term of Eq. (\ref{IEES}), using the gauge where $\xi=1$, acting with the differential operators and using the Fourier representation for the Dirac delta function $\delta^{4}\left(p-p'\right)=\int d^{4}x/\left(2\pi\right)^{4}e^{-ix\cdot(p-p')}$, we obtain
\begin{eqnarray}
\label{JJJJ1}
\int d^{4}x   \ \Bigl(\partial^{\alpha}\partial^{\beta}\Delta^{\nu\rho}\left(x,y\right)\Bigr)
\Delta^{\mu\gamma}\left(z,x\right) \nonumber\\
=-\eta^{\nu}_{\ \rho}\eta^{\mu}_{\ \gamma}\int\frac{d^{4}p}{(2\pi)^{4}}\frac{p^{\alpha}p^{\beta}}{p^{4}}\exp\left[-ip(z-y)\right] \ .
\end{eqnarray}
Now, using Eq's (\ref{laj}) and (\ref{JJJJ1}), the total interaction energy between external sources, up to the leading order in the background tensor, reads 
\begin{eqnarray}
\label{EF}
E&=&\frac{1}{2T}\int\int d^{4}y \ d^{4}z \ J^{\gamma}\left(z\right)D_{\gamma\rho}\left(z,y\right)J^{\rho}\left(y\right) \ ,
\end{eqnarray}
where we defined the propagator,
\begin{eqnarray}
\label{EFD}
D_{\gamma\rho}\left(z,y\right)&=&\int\frac{d^{4}p}{(2\pi)^{4}}\Biggl[-\frac{\eta_{\gamma\rho}}{p^{2}}
+K_{(F)\gamma\alpha\rho\beta}\frac{p^{\alpha}p^{\beta}}{p^{4}}\Biggr]\nonumber\\
&
&\times\exp\left[-ip\cdot(z-y)\right] \ .
\end{eqnarray}

From expression (\ref{EF}) we can compute the interaction energy between field sources up to first order in the background tensor $K_{(F)\gamma\alpha\rho\beta}$. It includes effects due to birefringent components as well as the nonbirefringent components of the background tensor. The exact propagator due to the CPT-even sector was studied in reference \cite{Fontes} for the very restrict case where the background tensor $K_{(F)\mu\nu\alpha\beta}$ was wrritten as a function of a single background vector $v^{\mu}=(v^{0},{\bf v})$, as follows  
\begin{equation}
\label{param}
K_{(F)\mu\nu\alpha\beta}=\eta_{\mu\alpha}v_{\nu}v_{\beta}-\eta_{\nu\alpha}v_{\mu}v_{\beta}+\eta_{\nu\beta}v_{\mu}v_{\alpha}-
\eta_{\mu\beta}v_{\nu}v_{\alpha} \ ,
\end{equation}
where it can be easily verified that this specific example satisfies the conditions (\ref{simetriasQ}) and (\ref{dtrace}).

\section{Point-like charges}
\label{III}

In this section we study the interaction energy between two stationary point-like charges following the same approach employed in Ref's \cite{BaroneHidalgo1,BaroneHidalgo2,Fontes,Fontes2}. The external sour\-ce which describes this system is given by
\begin{eqnarray}
\label{corre1Em}
J_{\rho}^{CC}({y})=q_{1}\eta^{0}_{\ \rho}\delta^{3}\left({\bf y}-{\bf a}_ {1}\right)+q_{2}\eta^{0}_{ \ \rho}\delta^{3}\left({\bf y}-{\bf a}_ {2}\right) \ ,
\end{eqnarray}
where we have two spatial Dirac delta functions concentrated at the positions ${\bf a}_{1}$ and ${\bf a}_{2}$. The parameters  $q_{1}$ and $q_{2}$ are the electric charges and the super-index $CC$ means that we have the interaction between two point-like charges.

Substituting Eq. (\ref{corre1Em}) in (\ref{EF}), discarding the self interaction contributions (the interactions of a given point-charge with itself), computing the integrals in the following order: $d^{3}{\bf y}$, $d^{3}{\bf z}$, $dy^{0}$, $dp^{0}$ and $dz^{0}$, using the fact that $\delta(p^{0})=\int dy^{0}/(2\pi)e^{ip^{0}y^{0}}$ and identifying the time interval as $T=\int dz^{0}$, we can write
\begin{eqnarray}
\label{EEI}
E^{CC}&=&q_{1}q_{2}\int\frac{d^{3}{\bf p}}{(2\pi)^{3}}\frac{\exp(i{\bf p}\cdot{\bf a})}{{\bf p}^2}-q_{1}q_{2}\nonumber\\
&
&\times\sum_{i,j=1}^{3} K_{(F)}^{ij}{\bf{\nabla}}_{{\bf a}}^{i}{\bf{\nabla}}_{{\bf a}}^{j}\int\frac{d^{3}{\bf p}}{(2\pi)^{3}}\frac{\exp(i{\bf p}\cdot{\bf a})}{{\bf p}^4},
\end{eqnarray}
where $i,j=1,2,3$ are spacial indexes and we defined the distance between the two electric charges ${\bf{a}={\bf a}_{2}-{\bf a}_{1}}=(a^{1},a^{2},a^{3})$ and the differential operator ${\bf {\nabla}}_{{\bf a}}^{i}=\partial/\partial a^{i}$.

The components $K_{(F)}^{ij}$ can be represented in a 3$\times$3 matrix, as follows
\begin{equation}
\label{matri11em}
K_{(F)}=\bordermatrix{&        \cr
							& K_{(F)}^{0101} \ \ & K_{(F)}^{0102} \ \ & K_{(F)}^{0103} \   \cr
              &  K_{(F)}^{0201} \ \ & K_{(F)}^{0202} \ \ & K_{(F)}^{0203} \       \cr
              &  K_{(F)}^{0301} \ \ & K_{(F)}^{0302} \ \ & K_{(F)}^{0303} \   \cr}\   \ .
\end{equation}

Using the fact that \cite{BaroneHidalgo1}
\begin{eqnarray}
\label{Ener4EM1}
\int\frac{d^{3}{\bf p}}{(2\pi)^{3}}\frac{\exp(i{\bf p}\cdot{\bf a})}{{\bf p}^2}=\frac{1}{4\pi |{\bf{a}}|} \ , \nonumber\\ 
\int\frac{d^{3}{\bf p}}{(2\pi)^{3}}\frac{\exp(i{\bf p}\cdot{\bf a})}{{\bf p}^4}=-\frac{|{\bf{a}}|}{8\pi} \ ,
\end{eqnarray}
and acting with the differentials operators, we obtain
\begin{eqnarray}
\label{maq}
E^{CC}=\frac{q_{1}q_{2}}{4\pi |{\bf{a}}|}\Bigl[1+\frac{1}{2}\Bigl(\sum_{i=1}^{3} K_{(F)}^{ii}-\sum_{i,j=1}^{3} K_{(F)}^{ij}\frac{a^{i}a^{j}}{{\bf a}^{2}}\Bigr)\Bigr] \ .
\end{eqnarray}

Eq. (\ref{maq}) gives us the interaction energy between two point-like charges for the model (\ref{modeloEM}). If we take $ K_{(F)}^{ij}=0$, the expression (\ref{maq}) reduces to the well-known Coulomb interaction. In Eq. (\ref{maq}) the summation $\sum_{i=1}^{3} K_{(F)}^{ii}$ is the trace of the matrix (\ref{matri11em}) which can be absorbed into the definition of the electric charges $q_{1}$ and $q_{2}$. The extra factor proportional to $\sum_{i,j=1}^{3} K_{(F)}^{ij}a^{i}a^{j}$ is an evident contribution which evinces the Lorentz symmetry breaking, leading us to an anisotropic interaction between the charges.

Taking $q_{1}=q_{2}=q$ and the limit ${\bf a}\to\infty$ in Eq. (\ref{maq}) we can show that the self energy of a point-like charge diverges, on the contrary to what happen in the nonminimal model considered in reference \cite{FBB2018} 

The interaction force between two charges can be obtained by taking the gradient of energy (\ref{maq}) with respect to ${\bf a}$, as follows
\begin{eqnarray}
\label{FI}
{\bf F}^{CC}&=&-{\bf {\nabla}}_{{\bf a}} E^{CC}\nonumber\\
&=&\frac{q_{1}q_{2}}{4\pi{\bf a}^{2}}\Biggl[\left(1+\frac{1}{2}\sum_{i=1}^{3} K_{(F)}^{ii}-\frac{3}{2}\sum_{i,j=1}^{3} K_{(F)}^{ij}\frac{a^{i}a^{j}}{{\bf a}^{2}}\right){{\hat a}}\nonumber\\
&
&+\frac{1}{|{\bf{a}}|}\sum_{i=1}^{3}a^{i}\left(K_{(F)}^{1i}{\hat{x}}+K_{(F)}^{2i}{\hat{y}}+K_{(F)}^{3i}{\hat{z}}\right)\Biggr] \ ,
\end{eqnarray}
where $\hat a$ is an unit vector pointing in the direction of the vector ${\bf a}$. In this paper $\hat{x},\hat{y},\hat{z}$ stand for the usual unit vectors in cartesian coordinates. 

The energy (\ref{maq}) exhibits an anisotropy due to the presence of the background tensor. As a consequence, we have the emergence of a spontaneous torque on an electric dipole, similarly to what was pointed out in Ref's \cite{Fontes,Fontes2}.

In order to investigate this effect, we consider a typical dipole composed by two opposite electric charges $q_{1}=-q_{2}=q$ placed at a fixed distance apart, at the positions ${\bf a}_{1}={\bf R}+\frac{{\bf d}}{2}$ and ${\bf a}_{2}={\bf R}-\frac{{\bf d}}{2}$, where $\bf d$ is taken to be a fixed vector. Let us also choose a simple ilustrative example for the background tensor which satisfies the properties (\ref{simetriasQ}) and (\ref{dtrace}), and cannot be expressed in the form (\ref{param}), as follows
\begin{eqnarray}
\label{parQ1}
 K_{(F)}^{0l0m}&=& 0 \ , \  
\mbox{for}\ \ \ l,m=1,2,3\ \ \mbox{and}\ l\neq m \ ,
\end{eqnarray} 
\begin{eqnarray}
\label{parQ2}
K_{(F)}^{0101}=0 \ , \ K_{(F)}^{0202}=-k_{1}\ , \  K_{(F)}^{0303}=k_{1} \ ,
\end{eqnarray} 
and
\begin{eqnarray}
\label{dtraceeee}
K^{lm}_{(F)\  lm}=0 \ , \mbox{for}\ \ \ l,m=1,2,3 \ .
\end{eqnarray}
where $k_{1}$ is a tiny constant.

In this specific case, the tensor (\ref{matri11em}) becomes 
\begin{equation}
\label{matri1em}
K_{(F)}=\bordermatrix{&        \cr
							& 0 \ \ &0 \ \ &0 \   \cr
              & 0 \ \ &-k_{1} \ \ &0 \       \cr
              & 0 \ \ &0 \ \ &k_{1} \   \cr}\   \ .
\end{equation}

Using Eq. (\ref{matri1em}), we can rewrite the energy (\ref{maq}) in the following way
\begin{eqnarray}
\label{maq2}
E^{CC}_{dipole}&=&-\frac{q^{2}}{4\pi |{\bf{d}}|}\Bigl[1-\frac{k_{1}}{2{\bf{d}}^{2}}
\Bigl(-\left({{\bf d}}\cdot\hat{y}\right)^{2}+\left({{\bf d}}\cdot\hat{z}\right)^{2}
\Bigr)\Bigr] \nonumber\\
&=&-\frac{q^{2}}{4\pi |{\bf{d}}|}\Bigl\{1-\frac{k_{1}}{2}\Bigl[1-\sin^{2}\theta\Bigl(1+\sin^{2}\phi\Bigr)\Bigr]\Bigr\} \ ,\nonumber\\
&\ &
\end{eqnarray}
where $0<\theta<\pi$ and $0<\phi<2\pi$ are the polar and azimuthal angles, in spherical coordinates (the $z$-axis is the polar axis), for the vector ${\bf{d}}$. 

From the energy (\ref{maq2}) have can compute two kinds of spontaneous torques acting on the vector ${\bf d}$, one related to the angle $\theta$ and another, to angle $\phi$, as follows 
\begin{eqnarray}
\label{TorqueEM}
\tau^{(\theta)}_{dipole}&=&-\frac{\partial E^{CC}_{dipole}}{\partial\theta}=\frac{q^{2}}
{8\pi |{\bf{d}}|}k_{1}\left(1+\sin^{2}\phi\right)\sin(2\theta) \ , \nonumber\\
\tau^{(\phi)}_{dipole}&=&-\frac{\partial E^{CC}_{dipole}}{\partial\phi}=\frac{q^{2}}
{8\pi |{\bf{d}}|}k_{1}\sin^{2}\theta\sin(2\phi) \ .
\end{eqnarray}

From expressions (\ref{TorqueEM}) we can see that $\tau^{(\theta)}_{dipole}=0$ when $\theta=0,\pi/2,\pi$, and for $\theta=\pi/4,\phi=\pi/2$, $\theta=\pi/4,\phi=3\pi/2$, $\tau^{(\theta)}_{dipole}$ attains its maximum intensity. For $\theta=0,\pi$ and $\phi=0,\pi/2,\pi,3\pi/2,2\pi$ the torque $\tau^{(\phi)}_{dipole}$ vanishes, for $\theta=\pi/2,\phi=\pi/4$ it has its maximum intensity.

It is convenient to rewrite our results in terms of the parameters defined in Refs. \cite{CPT1,CPT2}, which encloses the 19 components of the background tensor $K_{(F)\mu\nu\alpha\beta}$ in a parity-even and a parity-odd subsectors, represented by the matrices ${\tilde{\kappa}}_{e}$  and ${\tilde{\kappa}}_{o}$, respectively, as follows
\begin{eqnarray}
\label{componentsK1}
\left({\tilde{\kappa}}_{e+}\right)^{jk}&=&\frac{1}{2}\left(\kappa_{DE}+\kappa_{HB}\right)^{jk} \ ,\\
\label{componentsK33}
\left({\tilde{\kappa}}_{e-}\right)^{jk}&=&\frac{1}{2}\left(\kappa_{DE}-\kappa_{HB}\right)^{jk}-\delta^{jk}{\tilde{\kappa}}_{{\mathrm{tr}}} \ ,\\
\label{componentsK43}
{\tilde{\kappa}}_{{\mathrm{tr}}}&=&\frac{1}{3}{\mathrm{tr}}\left(\kappa_{DE}\right) \ ,\\
\label{componentsK53}
\left({\tilde{\kappa}}_{o+}\right)^{jk}&=&\frac{1}{2}\left(\kappa_{DB}+\kappa_{HE}\right)^{jk} \ ,\\
\label{componentsK63}
\left({\tilde{\kappa}}_{o-}\right)^{jk}&=&\frac{1}{2}\left(\kappa_{DB}-\kappa_{HE}\right)^{jk} \ ,
\end{eqnarray} 
where the $3\times3$ matrices $\kappa_{DE},\kappa_{HB},\kappa_{DB},\kappa_{HE}$ are defined in terms of the $K_{(F)}$-tensor components, as follows,
\begin{eqnarray}
\label{componentsK2}
\left(\kappa_{DE}\right)^{jk}&=&-2K_{(F)}^{0j0k}  , \left(\kappa_{HB}\right)^{jk}=\frac{1}{2}\epsilon^{jpq}\epsilon^{klm}K_{(F)}^{pqlm}, \nonumber\\
\left(\kappa_{DB}\right)^{jk}&=&-\left(\kappa_{HE}\right)^{kj}=\epsilon^{kpq}K_{(F)}^{0jpq} \ .
\end{eqnarray}

The matrices $\kappa_{DE}$ and $\kappa_{HB}$ contain together 11 independent components, while $\kappa_{DB}$ and $\kappa_{HE}$ contain together 8 components,
which sum the 19 independent elements of the background tensor $K_{(F)\mu\nu\alpha\beta}$. The 10 coefficients sensitive to birefringence are contained in the matrices ${\tilde{\kappa}}_{e+}$ and ${\tilde{\kappa}}_{o-}$, and the 9 nonbirefringent coefficients are contained in ${\tilde{\kappa}}_{e-}$ and ${\tilde{\kappa}}_{o+}$.  

The coeficient $k_{1}$ in Eq. (\ref{matri1em}) can be written as follows
\begin{eqnarray}
\label{componentsK3}
k_{1}=\frac{1}{2}\left[{(\tilde{\kappa}}_{e-})^{22}+({\tilde{\kappa}}_{e+})^{22}\right] \ .
\end{eqnarray}

Substituting (\ref{componentsK3}) in (\ref{maq2}), we arrive at
\begin{eqnarray}
\label{enerkec}
E^{CC}_{dipole}&=&-\frac{q^{2}}{4\pi |{\bf{d}}|}\Bigl\{1-\frac{1}{4}\left[{(\tilde{\kappa}}_{e-})^{22}+({\tilde{\kappa}}_{e+})^{22}\right]\nonumber\\
&
&\times\left[1-\sin^{2}\theta\left(1+\sin^{2}\phi\right)\right]\Bigr\} \ .
\end{eqnarray}

Substituting (\ref{componentsK3}) in (\ref{TorqueEM}), we obtain
\begin{eqnarray}
\label{TorqueEMke}
\tau^{(\theta)}_{dipole}&=&\frac{q^{2}}
{16\pi |{\bf{d}}|}\left[({\tilde{\kappa}}_{e-})^{22}+({\tilde{\kappa}}_{e+})^{22}\right]\cr
&\ &\times\left(1+\sin^{2}\phi\right)\sin(2\theta) \ , \nonumber\\
\tau^{(\phi)}_{dipole}&=&\frac{q^{2}}
{16\pi |{\bf{d}}|}\left[({\tilde{\kappa}}_{e-})^{22}+({\tilde{\kappa}}_{e+})^{22}\right]\cr
&\ &\times\sin^{2}\theta\sin(2\phi) \ .
\end{eqnarray}

From Eq. (\ref{TorqueEMke}) we can notice that the torques induced on an electric dipole, by the background tensor (\ref{matri1em}), are effects due to the parity-even components of the background tensor. These torques have contributions due to the nonbirefringent component, $({\tilde{\kappa}}_{e-})^{22}$, as well as contributions due to the birefringent component, $({\tilde{\kappa}}_{e+})^{22}$.

Another illustrative example which satisfies the conditions (\ref{simetriasQ}) and (\ref{dtrace}), and is not a particular case of (\ref{param}), is the following
\begin{eqnarray}
\label{paraQ3}
K_{(F)}^{0103}=k_{2}\ ,
\end{eqnarray}
with all other elements of the background tensor, $K_{(F)}$, taken to be equal to zero and where $k_{2}$ is a very tiny constant.

In this case, the matrix $K_{(F)}$ in (\ref{matri11em}) reads   
\begin{equation}
\label{matri120}
K_{(F)}=\bordermatrix{&        \cr
							& 0 \ \ &0 \ \ &k_{2} \   \cr
              & 0 \ \ &0 \ \ &0 \       \cr
              & k_{2} \ \ &0 \ \ &0 \   \cr}\   \ ,
\end{equation}
with the corresponding energy (\ref{maq}), 
\begin{eqnarray}
\label{PTorque}
E^{CC}_{dipole}=-\frac{q^{2}}{4\pi |{\bf{d}}|}\left(1-\frac{k_{2}}{2}\sin(2\theta)\cos \phi\right)\ .
\end{eqnarray}

In the same way, from the energy (\ref{PTorque}) we can obtain two kinds of spontaneous torques on the vector ${\bf d}$, as follows 
\begin{eqnarray}
\label{TorqueEMP}
\tau^{(\theta)}_{dipole}&=&-\frac{\partial E^{CC}_{dipole}}{\partial\theta}=-\frac{q^{2}}{4\pi |{\bf{d}}|}k_{2}{\cos (2\theta)}\cos\phi \ , \nonumber\\
\tau^{(\phi)}_{dipole}&=&-\frac{\partial E^{CC}_{dipole}}{\partial\phi}=\frac{q^{2}}{8\pi |{\bf{d}}|}k_{2}\sin(2\theta)\sin\phi \ .
\end{eqnarray}

If $\theta=\pi/4,3\pi/4$ and $\phi=\pi/2,3\pi/2$ the torque $\tau^{(\theta)}_{dipole}$ vanishes and for $\theta=0,\phi=0$, $\theta=\pi/2,\phi=\pi$ it exhibits its maximum intensity. If $\theta=0,\pi$ and $\phi=0,\pi,2\pi$ we have $\tau^{(\phi)}_{dipole}=0$ and for $\theta=\pi/4,\phi=\pi/2$, we have the maximum intensity for $\tau^{(\phi)}_{dipole}$.      

Using the Lorentz-breaking coefficients defined in Refs. \cite{CPT1,CPT2}, we can write
\begin{eqnarray}
\label{componentsK333}
k_{2}=-\frac{1}{2}\left[{(\tilde{\kappa}}_{e-})^{31}+({\tilde{\kappa}}_{e+})^{31}\right] \ ,
\end{eqnarray}
So, substituting (\ref{componentsK333}) in (\ref{PTorque}), we obtain
\begin{eqnarray}
\label{PTorque2}
E^{CC}_{dipole}&=&-\frac{q^{2}}{4\pi |{\bf{d}}|}\Bigl\{1+\frac{1}{4}\left[{(\tilde{\kappa}}_{e-})^{31}+({\tilde{\kappa}}_{e+})^{31}\right]\nonumber\\
&
&\times\sin(2\theta)\cos \phi\Bigr\}\ .
\end{eqnarray}
and the torques (\ref{TorqueEMP}) become
\begin{eqnarray}
\label{TorqueEMPke}
\tau^{(\theta)}_{dipole}&=&\frac{q^{2}}{8\pi |{\bf{d}}|}\left[{(\tilde{\kappa}}_{e-})^{31}+({\tilde{\kappa}}_{e+})^{31}\right]{\cos (2\theta)}\cos\phi \ , \nonumber\\
\tau^{(\phi)}_{dipole}&=&-\frac{q^{2}}{16\pi |{\bf{d}}|}\left[{(\tilde{\kappa}}_{e-})^{31}+({\tilde{\kappa}}_{e+})^{31}\right]\sin(2\theta)\sin\phi \ . \nonumber\\
&\ &
\end{eqnarray}

Once again, we have an example where the expontaneous torques are induced by the parity-even sector of the background tensor, with contributions of its birefringent and nonbirefringent component, $({\tilde{\kappa}}_{e+})^{31}$ and $({\tilde{\kappa}}_{e-})^{31}$, respectively, in this case

If we use the tensor (\ref{param}) in the Lagrangian (\ref{modeloEM}), it becomes equivalent to the one considered in Ref. \cite{Fontes}. Substituting Eq. (\ref{param}) in (\ref{maq}) and (\ref{FI}), we obtain
\begin{eqnarray}
\label{ENB}
E^{CC}(v)&=&\frac{q_{1}q_{2}}{4\pi |{\bf{a}}|}\Biggl[1-(v^{0})^{2}+\frac{1}{2}\Biggl({\bf v}^{2}-\frac{({\bf v}\cdot{\bf a})^{2}}{{\bf a}^{2}}\Biggr)\Biggr], \\
\label{ENBII5}
{\bf F}^{CC}(v)&=&\frac{q_{1}q_{2}}{4\pi{\bf a}^{2}}\Biggl[\Biggl(1-(v^{0})^{2}+\frac{1}{2}{\bf v}^{2}\Biggr){\hat a}\nonumber\\
&
&+({\bf v}\cdot{\hat a})\Biggl({\bf v}-\frac{3}{2}({\bf v}\cdot{\hat a}){\hat a}\Biggr)\Biggr]\ ,
\end{eqnarray}
which are in perfect agreement with the result of reference \cite{Fontes} if they are expanded up to order of $v^{2}$. 

To put the possible signals of Lorentz-symmetry breaking  obtained in this section in a physical context and obtain estimates in order of magnitude for the Lorentz-symmetry breaking parameters involved, let us consider experimental data for the hydrogen atom, which is a system whose dynamics is governed by the Coulomb (electrostatic) potential. This approach is justified because any deviation from the Coulomb potential would bring out experimental signals on hydrogen atoms. The ground state energy for the hydrogen atom has an uncertainty given by $~6.1\times10^{-9}$ \cite{dadosQED}. From Eq. (\ref{maq}) we can see that the relative correction in the Coulomb behavior imposed by the Lorentz symmetry breaking is proportional to the coefficients $K_{(F)}^{ij}$, or equivalently, the ones defined in Eqs. (\ref{componentsK1}), (\ref{componentsK33}), (\ref{componentsK43}), (\ref{componentsK53}), and (\ref{componentsK63}). So, by using atomic data, we can overestimate an upperbound for the coefficients (\ref{componentsK1}), (\ref{componentsK33}), (\ref{componentsK43}), (\ref{componentsK53}), (\ref{componentsK63}) as being of order $~6.1\times10^{-9}$. We can search for more restrictive estimates and use the hyperfine corrections to the hydrogen spectra, which are proportional to the fine structure constant, whose relative uncertainty is $~2.3\times10^{-10}$ \cite{dadosQED2}. So this more restictive estimates predicts an upper bound of order $~2.3\times10^{-10}$ for (\ref{componentsK1}), (\ref{componentsK33}), (\ref{componentsK43}), (\ref{componentsK53}) and (\ref{componentsK63}). These estimates are far beyond the ones obtained from optical methods \cite{dados}.

In order to investigate if the torques (\ref{TorqueEMke}) and (\ref{TorqueEMPke}) can be of some relevance in condesend matter physics, let us consider a typical microscopic system of condensend matter, with distances in order of angstrons ($|{\bf d}|\sim\mbox{\AA})$, electric charges equal, in magnitude, to the electron's one ($q^{2}\sim 2.899\times10^{-27}Nm^{2}$) and the overestimated values for the Lorentz-symmetry breaking parameters obtained from Ref. \cite{dados} (${(\tilde{\kappa}}_{e-})^{ij}\sim4\times10^{-18}$, ${(\tilde{\kappa}}_{e+})^{ij}\sim2\times10^{-37}$). In this case, we have for the torques (\ref{TorqueEMke}) and (\ref{TorqueEMPke}), $\tau\sim10^{-36}Nm$. This very small result suggests that these kinds of effects are out of reach of measurements nowadays in systems of condensed matter.

\section{A steady line current and a point-like charge}
\label{IV}
In this Section we study the interaction energy between a steady line current and a point-like stationary charge. Such interaction does not exist in the Maxwell electrodynamics, as discussed in references \cite{Fontes,Fontes2}. The steady line current shall be taken to flow parallel to the $z$-axis, along the straight line located at ${\bf A}=(A^{1},A^{2},0)$. The electric charge is placed at position ${\bf u}$. The external source for this system is given by
\begin{eqnarray}
\label{corre3Em}
J_{\rho}^{SC}\left(y\right)=I\eta^{3}_{\ \rho}\delta^{2}\left({\bf y}_{\perp}-{\bf A}\right)+q\eta^{0}_{\ \rho}\delta^{3}\left({\bf y}-{\bf u}\right) \ .
\end{eqnarray}
where ${\bf y}_{\perp}=(y^{1},y^{2},0)$, is the position vector perpendicular to the straight line current. The parameters $I$ and $q$ stand for, respectively, the current intensity and electric charge strength. The super-index $SC$ means that we have a system composed by a steady line current and a point-like charge.  

Substituting Eq. (\ref{corre3Em}) in (\ref{EF}), discarding self-interac-tion terms, performing the integrals in the following order: $d^{2}{\bf y_{\perp}}$, $d^{2}{\bf z_{\perp}}$, $dz^{3}$, $dy^{3}$, $dp^{3}$, $dy^{0}$, $dp^{0}$ and $dz^{0}$ and identifying the time interval $\int dz^{0}=T$, we obtain
\begin{eqnarray}
\label{EE2}
E^{SC}=-qI\sum_{i,j=1}^{2}K_ {(F)\perp}^{ij}{\bf{\nabla}}_{{\bf a}_{\perp}}^{i}{\bf{\nabla}}_{{\bf a}_{\perp}}^{j}\int\frac{d^{2}{\bf p}_{\perp}}{(2\pi)^{2}}\frac{e^{i{\bf p}_{\perp}\cdot{\bf a}_{\perp}}}{{\bf p}_{\perp}^4}\ ,
\end{eqnarray}
where ${\bf p}_{\perp}=(p^{1},p^{2},0)$ and ${\bf a}_{\perp}={\bf A}-{\bf u}=(A^{1}-u^{1},A^{2}-u^{2},0)=(a^{1},a^{2},0)$ is the distance between the charge and the line current. We also defined the differential operator ${\bf{\nabla}}_{{\bf a}_{\perp}}^{i}=\partial/\partial a^{i}$ (with $i=1,2$) and the 2$\times$2 matrix $K_{(F)\perp}^{ij}$ as follows 
\begin{equation}
\label{matri1em2}
K_{(F)\perp}=\bordermatrix{&        \cr
							& K_{(F)}^{0131} \ \ &K_{(F)}^{0132} \  \cr
              & K_{(F)}^{0231} \ \ &K_{(F)}^{0232}  \    \cr}\   \ .
\end{equation}

The integral in Eq. (\ref{EE2}) is divergent. In order to solve this problem we proceed as in reference \cite{BaroneHidalgo1,Fontes}, introducing a regulator parameter with mass dimension, as follows
\begin{eqnarray}
\label{EE3}
E^{SC}&=&-qI\sum_{i,j=1}^{2}K_{(F)\perp}^{ij}\nonumber\\
 & &\times{\bf{\nabla}}_{{\bf a}_{\perp}}^{i}{\bf{\nabla}}_{{\bf a}_{\perp}}^{j}\lim_{m\rightarrow 0}\int\frac{d^{2}{\bf p}_{\perp}}{(2\pi)^{2}}\frac{e^{i{\bf p}_{\perp}\cdot{\bf a}_{\perp}}}{({\bf p}_{\perp}^2+m^{2})^{2}} \nonumber \\
&=&qI\sum_{i,j=1}^{2}K_{(F)\perp}^{ij}{\bf{\nabla}}_{{\bf a}_{\perp}}^{i}{\bf{\nabla}}_{{\bf a}_{\perp}}^{j}\nonumber\\
&
&\times\lim_{m\rightarrow 0}\frac{1}{2m}\frac{\partial}{\partial m}\int\frac{d^{2}{\bf p}_{\perp}}{(2\pi)^{2}}\frac{e^{i{\bf p}_{\perp}\cdot{\bf a}_{\perp}}}{{\bf p}_{\perp}^2+m^{2}} \ .
\end{eqnarray}
Using the fact that \cite{BaroneHidalgo1}
\begin{eqnarray}
\label{int4EM}
\int\frac{d^{2}{\bf q}_{\perp}}{(2\pi)^{2}}\frac{\exp(i{\bf p}_{\perp}\cdot{\bf a}_{\perp})}{{\bf p}_{\perp}^2+m^{2}}=\frac{1}{2\pi}K_{0}(m|{{\bf a}}_{\perp}|) \ ,
\end{eqnarray}
and acting with the differential operators, we arrive at
\begin{eqnarray}
\label{EE4}
E^{SC}&=&-\frac{qI}{4\pi}\Bigl\{\lim_{m\rightarrow 0}[-K_{0}(m|{{\bf a}}_{\perp}|)]\sum_{i=1}^{2}K_ {(F)\perp}^{ii}\nonumber\\
&
&+\lim_{m\rightarrow 0}mK_{1}(m|{{\bf a}}_{\perp}|)\sum_{i,j=1}^{2}K_{(F)\perp}^{ij}\frac{a^{i}a^{j}}{|{{\bf a}}_{\perp}|}\Bigr\} \ ,
\end{eqnarray}
where $K_{0}(m|{{\bf a}}_{\perp}|)$ and $K_{1}(m|{{\bf a}}_{\perp}|)$ stand for the K-Bessel functions.

Using the fact that \cite{Arfken} 
\begin{eqnarray}
\label{kbessel}
-K_{0}(m|{{\bf a}}_{\perp}|)&\stackrel{m\rightarrow0}{\rightarrow}&\ln\left(m|{{\bf a}}_{\perp}|/2\right)+\gamma \nonumber\\
 mK_{1}(m|{{\bf a}}_{\perp}|)&\stackrel{m\rightarrow0}{\rightarrow}&1/|{{\bf a}}_{\perp}|, 
\end{eqnarray}
where $\gamma$ is the Euler constant, we rewrite Eq. (\ref{EE4}) in the following form
\begin{eqnarray}
\label{EE5}
E^{SC}&=&-\frac{qI}{4\pi}\Biggl\{\lim_{m\rightarrow 0}\left[\ln\left(\frac{m|{{\bf a}}_{\perp}|}{2}\right)+\gamma\right]\sum_{i=1}^{2}K_ {(F)\perp}^{ii}\nonumber\\
&
&+\sum_{i,j=1}^{2}K_{(F)\perp}^{ij}\frac{a^{i}a^{j}}{{\bf a}_{\perp}^{2}}\Biggr\}\nonumber\\ 
&=&-\frac{qI}{4\pi}\Biggl\{\lim_{m\rightarrow 0}\Bigl[\ln\left(\frac{m|{{\bf a}}_{\perp}|}{2}\right)+\gamma+\ln(ma_{0})\nonumber\\
&
&-\ln(ma_{0})\Bigr]\sum_{i=1}^{2}K_ {(F)\perp}^{ii}+\sum_{i,j=1}^{2}K_{(F)\perp}^{ij}\frac{a^{i}a^{j}}{{\bf a}_{\perp}^{2}}\Biggr\} \nonumber\\
&=&-\frac{qI}{4\pi}\Biggl\{\Bigl[\ln\left(\frac{|{{\bf a}}_{\perp}|}{a_{0}}\right)+\gamma-\ln 2+\lim_{m\rightarrow 0}\ln(ma_{0})\Bigr]\nonumber\\
&
&\times\sum_{i=1}^{2}K_ {(F)\perp}^{ii}+\sum_{i,j=1}^{2}K_{(F)\perp}^{ij}\frac{a^{i}a^{j}}{{\bf a}_{\perp}^{2}}\Biggr\}\ ,
\end{eqnarray}
where, in the third line of Eq. (\ref{EE5}), we added and subtract the quantity $\ln(ma_{0})$, where $a_{0}$ is an arbitrary length-dimensional constant. Neglecting, in the penultimate line of Eq. (\ref{EE5}), the terms that do not depend on the distance ${\bf a}_{\perp}$, since they do not contribute to the force between the line current and the point charge, we obtain
\begin{eqnarray}
\label{EE6}
E^{SC}&=&-\frac{qI}{4\pi}\Bigg[\ln\left(\frac{|{{\bf a}}_{\perp}|}{a_{0}}\right)\sum_{i=1}^{2}K_{(F)\perp}^{ii}\cr\cr
&\ &+\sum_{i,j=1}^{2}K_{(F)\perp}^{ij}\frac{a^{i}a^{j}}{{\bf a}_{\perp}^{2}}\Bigg].
\end{eqnarray}

The interaction energy (\ref{EE6}) is an effect due solely to the Lorentz symmetry breaking. In Eq. (\ref{EE6}), the coefficient $\sum_{i=1}^{2}K_{(F)\perp}^{ii}$ is the trace of the matrix (\ref{matri1em2}), which can be absorbed into the definition of the current intensity $I$, and electric charge $q$. The term proportional to $\sum_{i,j=1}^{2}K_{(F)\perp}^{ij}a^{i}a^{j}$ is an evident contribution which evinces the Lorentz-symmetry breaking, leading us to an anisotropic interaction between the line current and the point-like charge.

The force on the point charge can be obtained from Eq. (\ref{EE6}) as follows,   
\begin{eqnarray}
\label{FI2}
{\bf F}^{SC}&=&-{\bf {\nabla}}_{{\bf a}_{\perp}}E^{SC}\nonumber\\
&=&-\frac{qI}{4\pi |{{\bf a}}_{\perp}|}\Biggl[\Biggl(\sum_{i=1}^{2}K_ {(F)\perp}^{ii}-2\sum_{i,j=1}^{2}K_{(F)\perp}^{ij}\frac{a^{i}a^{j}}{{\bf a}_{\perp}^{2}}\Biggr){\hat{a}_{\perp}}\nonumber\\
&\ &+\frac{1}{|{{\bf a}}_{\perp}|}\sum_{i=1}^{2}{a_{\perp}^{i}}
\Bigl[\Bigl(K_{(F)\perp}^{i1}+K_{(F)\perp}^{1i}\Bigr){\hat{x}}\nonumber\\
&\ &+\left(K_{(F)\perp}^{i2}+K_{(F)\perp}^{2i}\right){\hat{y}}\Bigr]\Biggr] \ ,
\end{eqnarray}
where ${\hat{a}_{\perp}}$ is the unit vector pointing on the direction of ${{\bf a}}_{\perp}$ and we used the fact that
\begin{equation} 
\label{aperp}
{\bf {\nabla}}_{{\bf a}_{\perp}}=\left(\frac{\partial}{\partial a^{1}},\frac{\partial}{\partial a^{2}},0\right)\ .
\end{equation}

The second term inside brackets in the interaction energy (\ref{EE6}) leads to a torque on the steady line current when we fix the point-like charge. In order to calculate this torque in a specific example (which is in accordance with the properties (\ref{simetriasQ}) and (\ref{dtrace})), let us take
\begin{eqnarray}  
\label{paraQ6}
K_{(F)}^{0131}=K_{(F)}^{0132}=K_{(F)}^{0231}=0 \ , \ K_{(F)}^{0232}=k_{3} \ ,
\end{eqnarray}
where $k_{3}$ is a tiny constant. 

In this case, the matrix (\ref{matri1em2}) reads 
\begin{equation}
\label{matri1em3}
K_{(F)\perp}=\bordermatrix{&        \cr
							& 0 \ \ &0 \  \cr
              & 0 \ \ &k_{3}  \    \cr}\   \ .
\end{equation}

Substituting Eq. $(\ref{matri1em3})$ in $(\ref{EE6})$, we obtain  
\begin{eqnarray}
\label{EE65}
E^{SC}&=&-\frac{qI}{4\pi}k_{3}\left[\ln\left(\frac{|{{\bf a}}_{\perp}|}
{a_{0}}\right)+\frac{\left({{\bf a}}_{\perp}\cdot{\hat{y}}\right)^{2}}
{{{\bf a}}_{\perp}^{2}}\right]\nonumber\\
&=&-\frac{qI}{4\pi}k_{3}\left[\ln\left(\frac{|{{\bf a}}_{\perp}|}
{a_{0}}\right)+\sin^{2}\alpha\right] \ ,
\end{eqnarray}
where $\alpha$ is the angle between the vector ${{\bf a}}_{\perp}$ and the unit vector ${\hat x}$ (the angle of the vector ${{\bf a}}_{\perp}$ in polar coordinates).

If we take a setup where the distance between the current line and the point-charge is fixed, the energy (\ref{EE65}) leads us to a torque in the whole system, as follows
\begin{eqnarray}
\label{TorqueEMII}
\tau^{SC}=-\frac{\partial E^{SC}}{\partial\alpha}=\frac{qI}{4\pi}k_{3}\sin(2\alpha) \ .
\end{eqnarray}
Notice that for $\alpha=0,\pi/2,\pi,2\pi$ the torque vanishes and for $\alpha=\pi/4$ and $\alpha=3\pi/4$ we have its maximum intensity.

The constant $k_{3}$ in (\ref{matri1em3})   can be written in terms of the Lorentz-breaking coefficients defined in Ref's \cite{CPT1,CPT2}, as follows
\begin{equation}
\label{k3}
k_{3}=-\frac{1}{2}\left[(\tilde{\kappa}_{o+})^{21}+(\tilde{\kappa}_{o-})^{21}\right] \ .
\end{equation}

In this way, the energy (\ref{EE65}) and the torque (\ref{TorqueEMII}) become
\begin{eqnarray}
\label{EMIIko}
E^{SC}&=&\frac{qI}{8\pi}\left[(\tilde{\kappa}_{o+})^{21}+(\tilde{\kappa}_{o-})^{21}\right]\nonumber\\
&
&\times\left[\ln\left(\frac{|{{\bf a}}_{\perp}|}
{a_{0}}\right)+\sin^{2}\alpha\right] \ .\\
\label{TorqueEMIIko}
\tau^{SC}&=&-\frac{qI}{8\pi}\left[(\tilde{\kappa}_{o+})^{21}+(\tilde{\kappa}_{o-})^{21}\right]\sin(2\alpha) \ .
\end{eqnarray}

From the result (\ref{TorqueEMIIko}) we can notice that the torque acting on the current line, for the specific example (\ref{matri1em3}), is an effect due to the parity-odd components of the background tensor. This torque exhibits contributions comming from the nonbirefringent component $({\tilde{\kappa}}_{o+})^{21}$ and from the birefringent component $({\tilde{\kappa}}_{o-})^{21}$ of the background tensor.

An infinite straight line current is an idealization. In a more realistic situation, one might take into account a finite length for the line current. The results obtained in this section must be considered when edge effects are negligible, what happens when the length of the line current is much higher in comparison with the distance between the line current and the point-charge.

If we substitute Eq. (\ref{param}) in (\ref{EE6}), we obtain
\begin{eqnarray}
\label{ENBII}
E^{SC}\left(v\right)=\frac{qI}{2\pi}v^{0}v^{3}\ln\left(\frac{|{{\bf a}}_{\perp}|}{a_{0}}\right)+\frac{qI}{4\pi}v^{0}v^{3} \ .
\end{eqnarray}
The second term in the right-hand side of Eq. (\ref{ENBII}) does not depend of distance $|{\bf a}_{\perp}|$, so it can be neglected. Therefore
\begin{eqnarray}
\label{ENBIII}
E^{SC}\left(v\right)=\frac{qI}{2\pi}({{\bf v}\cdot{\hat z}})\ v^{0}\ln\left(\frac{|{{\bf a}}_{\perp}|}{a_{0}}\right) \ ,
\end{eqnarray}
where $v^{3}={{\bf v}\cdot{\hat z}}$ is the projection of the vector ${\bf v}$ along the straight line current. This result is in accordance with the reference \cite{Fontes}, up to order of $v^{2}$.

Once again, let us investigate if the above effects can have some relevance in condensed matter systems. The highest electric currents achieved in laboratory are of magnitude $10^{5}A$, the overestimated values from Ref. \cite{dados}, $(\tilde{\kappa}_{o+})^{ij}\sim 2\times10^{-18}$, $(\tilde{\kappa}_{o-})^{ij}\sim2\times10^{-37}$, and the charge of the electron, we have the estimate value for the torque (\ref{TorqueEMIIko}), $\tau^{SC}\sim10^{-30}Nm$.

In the opposite direction, we can search for some Lorentz-symmetry breaking signals in current jets produced on galaxies, where we find the highests electric currents in nature \cite{AstrJour}, of magnitude $\sim10^{18}A$. In this case, the torque is estimated by $\tau^{SC}\sim10^{-17}Nm$. These small result suggests that the effects of the torque (\ref{TorqueEMIIko}) are far beyond any measurement range nowadays.

\section{Dirac strings}
\label{V}

In this section we consider, firstly, a system composed by a point-like charge placed at position ${\bf a}$ and a Dirac string. This system is described by the external source
\begin{eqnarray}
\label{Dcurrent1}
J_{\rho}^{DC}\left(y\right)=J_{(D)\rho}\left(y\right)+q\eta_{\ \rho}^{0}\delta^{3}({\bf y}-{\bf a}) \ ,
\end{eqnarray}
where $J_{(D)}^{\rho}\left(y\right)$ stands for the external field source produced by the Dirac string and the second term on the right hand side is the source produced by the point-like charge. The super-index $DC$ means that we have a Dirac string and a point-like charge.

Now, we choose a coordinate system where the Dirac string lies along the $z$-axis with internal magnetic flux $\Phi$. Its corresponding source is given by \cite{Fontes,Fontes2,MBB,FernandaDissertacao,AndersonDissertacao}
\begin{equation}
\label{Dircurr2}
J_{(D)}^{\rho}({y})=i\Phi(2\pi)^{2}\int\frac{d^{4}p}{(2\pi)^{4}}\delta(p^{0})\delta(p^{3})\varepsilon^{0\rho}_{\ \ \nu3}\ p^{\nu}e^{-ipy}\ ,
\end{equation}
where $\varepsilon^{\mu\nu\alpha\beta}$ stands for Levi-Civita tensor with $\varepsilon^{0123}=1$. If $\Phi>0$, we have the internal magnetic field along $\hat z$. For $\Phi<0$, the internal magnetic field points in the opposite direction. 

From now on, in this Section, the sub-index $\perp$ means that we are taking just the components of a given vector perpendicular to the string. For instance, ${\bf p}_{\perp}=(p^{1},p^{2},0)$ is the momentum perpendicular to the string.

Substituting Eq. (\ref{Dcurrent1}) in (\ref{EF}), discarding self-interac\-tion terms, which do not contribute to the force between the string and the charge (the self-interaction terms are proportional to $q^{2}$ or $\Phi^{2}$ ) and following similar steps employed in the previous sections, we can show that
\begin{eqnarray}
\label{EE10}
&&E^{DC}=-\frac{q\Phi}{4\pi |{{\bf a}}_{\perp}|}\Bigl[\left({{\bf a}_{\perp}}\cdot{\hat{x}}\right)K_{(F)}^{0121}-\left({{\bf a}_{\perp}}\cdot{\hat{y}}\right)K_{(F)}^{0212}\Bigr]\nonumber\\
&
&\times\lim_{m\rightarrow 0}\Bigl[2mK_{1}(m|{{\bf a}}_{\perp}|)-|{{\bf a}}_{\perp}|m^{2}K_{0}(m|{{\bf a}}_{\perp}|) \Bigr]\ .
\end{eqnarray}
Now, taking the limit $m\rightarrow0$, we arrive at 
\begin{eqnarray}
\label{EE100}
E^{DC}&=&-\frac{q\Phi}{2\pi{\bf a}_{\perp}^{2}}\Bigl[\left({{\bf a}_{\perp}}\cdot{\hat{x}}\right)K_{(F)}^{0121}-\left({{\bf a}_{\perp}}\cdot{\hat{y}}\right)K_{(F)}^{0212}\Bigr] \ .
\end{eqnarray}

We notice that Eq. (\ref{EE100}) is also an effect due solely to the Lorentz symmetry breaking.

The energy (\ref{EE100}) leads to a force between the Dirac string and the charge as well as a torque on the string, if we take a setup where the distance between the charge and the string is fixed. Defining the tiny constants $k_{4}$ and $k_{5}$,
\begin{eqnarray}
\label{paraQ7}
K_{(F)}^{0112}=k_{4} \ , \ \ K_{(F)}^{0212}=k_{5}\ , 
\end{eqnarray}
we can write Eq's (\ref{paraQ7}) in (\ref{EE100}) in the form
\begin{eqnarray}
\label{EE1011}
E^{DC}&=&\frac{q\Phi}{2\pi{\bf{a}}_{\perp}^{2}}\left[k_{4}
\left({\bf{a}}_{\perp}\cdot{\hat{x}}\right)+k_{5}
\left({\bf{a}}_{\perp}\cdot{\hat{y}}\right)\right]\nonumber\\
&=&\frac{q\Phi}{2\pi\mid{\bf{a}}_{\perp}\mid}
\left(k_{4}\cos\alpha+ k_{5}\sin\alpha\right)\ ,
\end{eqnarray}
what leads to a torque on the Dirac string, as follows 
\begin{eqnarray}
\label{tostring}
\tau^{DC}=-\frac{\partial E^{DC}}{\partial\alpha}=
\frac{q\Phi}{2\pi\mid{\bf{a}}_{\perp}\mid}\left(k_{4}\sin\alpha
-k_{5}\cos\alpha\right)\ .
\end{eqnarray}
where $\alpha$ is the angle between ${\bf a}_{\perp}$ and ${\hat x}$.

In terms of the Lorentz-breaking  coefficients defined in Ref's \cite{CPT1,CPT2}, we have 
\begin{eqnarray}
\label{k4k5}
k_{4}=\frac{1}{2}\left[(\tilde{\kappa}_{o+})^{13}+(\tilde{\kappa}_{o-})^{13}\right] \ ,\nonumber\\
k_{5}=\frac{1}{2}\left[(\tilde{\kappa}_{o+})^{23}+(\tilde{\kappa}_{o-})^{23}\right] \ .
\end{eqnarray}

From the Eqs. (\ref{EE1011}), (\ref{tostring}) and (\ref{k4k5}), we can write
\begin{eqnarray}
E^{DC}&=&\frac{q\Phi}{4\pi\mid{\bf{a}}_{\perp}\mid}
\Bigl\{\left[(\tilde{\kappa}_{o+})^{13}+(\tilde{\kappa}_{o-})^{13}\right]\cos\alpha\nonumber\\
&
&+\left[(\tilde{\kappa}_{o+})^{23}+(\tilde{\kappa}_{o-})^{23}\right]\sin\alpha\Bigr\} \ .\\
\tau^{DC}&=&
\frac{q\Phi}{4\pi\mid{\bf{a}}_{\perp}\mid}\Bigl\{\left[(\tilde{\kappa}_{o+})^{13}\sin\alpha
-(\tilde{\kappa}_{o+})^{23}\cos\alpha\right]\nonumber\\
&
&+\left[(\tilde{\kappa}_{o-})^{13}\sin\alpha
-(\tilde{\kappa}_{o-})^{23}\cos\alpha\right]\Bigr\}\ .
\label{tostringko}
\end{eqnarray}

The torque in Eq. (\ref{tostringko}) is an effect due to the parity-odd components of the background tensor. The first contribution inside brackets on the right hand side of the Eq. (\ref{tostringko}) comes from the nonbirefringent sector of the background tensor and the second term comes from the birefringent sector.

Just for completeness, we consider the interaction between a Dirac string and a steady line current, both parallel to each other. The corresponding external source is given by
\begin{equation}
J^{DS}_{\rho}({y})=J_{\rho(D)}\left({y}\right)+I\eta^{3}_{\ \rho}\delta^{2}\left({\bf y}_{\perp}-{\bf a}_{\perp}\right)
\end{equation}
where $J_{(D)}^{\rho}\left({y}\right)$ is given by (\ref{Dircurr2}). The super-index $DS$ means that we have a Dirac string and a steady line current.

Following similar steps employed previously, we obtain the result
\begin{eqnarray}
\label{EE105}
{\cal E}^{DS}&=&\frac{E^{DS}}{L}\cr\cr
&=&-\frac{I\Phi}{2\pi{\bf a}_{\perp}^{2}}\Bigl[\left({{\bf a}_{\perp}}\cdot{\hat{x}}\right)K_{(F)}^{3121}-\left({{\bf a}_{\perp}}\cdot{\hat{y}}\right)K_{(F)}^{3212}\Bigr]\ ,
\end{eqnarray}
where we identified the length of the Dirac string, $L=\int dz^{3}$, defined the energy per unit of string length ${\cal E}$.

From the energy (\ref{EE105}), which is an effect due solely to the Lorentz symmetry breaking, we can obtain a force between the Dirac string and the steady line current, as well as a torque between them.

Considering $\alpha$ the angle between ${\bf a}_{\perp}$ and ${\hat x}$, and defining
\begin{equation}
\label{k6k7}
K_{(F)}^{3121}=k_{6} \ , \ \ \ K_{(F)}^{3212}=k_{7} \ ,
\end{equation}
the energy (\ref{EE105}) reads
\begin{equation}
\label{eneralpha}
{\cal E}^{DS}=-\frac{I\Phi}{2\pi\mid{\bf a}_{\perp}\mid}\left(k_{6}\cos\alpha-k_{7}\sin\alpha\right)\ ,
\end{equation}
where $k_{6}$ and $k_{7}$ are tiny constants.

Using the Lorentz-breaking coefficients defined in Ref's \cite{CPT1,CPT2}, we obtain
\begin{eqnarray}
\label{k6k7energy67}
k_{6}=\frac{1}{2}\left[(\tilde{\kappa}_{e-})^{23}-(\tilde{\kappa}_{e+})^{23}\right] \ ,\nonumber\\
k_{7}=\frac{1}{2}\left[(\tilde{\kappa}_{e-})^{13}-(\tilde{\kappa}_{e+})^{13}\right] \ ,
\end{eqnarray}
and the energy (\ref{eneralpha}) becomes
\begin{eqnarray}
\label{lllll}
{\cal E}^{DS}&=&-\frac{I\Phi}{4\pi\mid{\bf{a}}_{\perp}\mid}\Bigl\{\left[-(\tilde{\kappa}_{e+})^{23}\cos\alpha
+(\tilde{\kappa}_{e+})^{13}\sin\alpha\right]\nonumber\\
&
&+\left[(\tilde{\kappa}_{e-})^{23}\cos\alpha
-(\tilde{\kappa}_{e-})^{13}\sin\alpha\right]\Bigr\} \ .
\end{eqnarray}

The energy in Eq. (\ref{lllll}) is an effect due to the parity-even components of the background tensor. The first contribution inside brackets on the right hand side of the Eq. (\ref{lllll}) is birefringent and the second one is nonbirefringent.


With the specific example (\ref{param}), Eq. (\ref{EE105}) becomes  
\begin{equation}
\label{EIIVNB}
{\cal E}^{DS}\left(v\right)=\frac{I\Phi}{2\pi{\bf a}_{\perp}^{2}}({{\bf v}\cdot{\hat z}})
\left[{\hat z}\cdot\left({\bf a}_{\perp}\times{\bf v}_{\perp}\right)\right] \ ,
\end{equation}
which is in accordance with \cite{Fontes}.

It is worth mentioning that an infinite Dirac string is an idealization of an infinitely long solenoid with vanishing radius and finite internal magnetic flux. In a more realistic situation, one might take into account a non-vanishing radius and a finite length for the solenoid. The results obtained in this section must be considered when the length of the solenoid is much higher than the distance between the line current and the point-charge and when the radius of the solenoid is negligible in comparison with this distance.

\section{Aharonov-Bohm bound states}
\label{Aharonov}

In this section we consider a simplified version of the so called Aharonov-Bohm bound states in the scenario described by the model (\ref{modeloEM}). For this task we start by considering the propagator (\ref{EFD}) and the field configuration produced by a given external source,
\begin{eqnarray}
\label{fields}
A^{\mu}\left(x\right)&=&\int d^{4}y \ D^{\mu\nu}\left(x,y\right)J_{\nu}\left(y\right)\nonumber\\
&=&\int d^{4}y\Bigl[D_{M}^{\mu\nu}\left(x,y\right)+D_{LV}^{\mu\nu}\left(x,y\right)\Bigr]J_{\nu}\left(y\right) ,
\end{eqnarray}
where we defined the standard Maxwell propagator,
\begin{eqnarray}
\label{propm}
D_{M}^{\mu\nu}\left(x,y\right)=-\int\frac{d^{4}p}{(2\pi)^{4}}\frac{\eta^{\mu\nu}}{p^{2}}e^{-ip\cdot(x-y)} \ ,
\end{eqnarray}
and the correction
\begin{eqnarray}
\label{proplv}
D_{LV}^{\mu\nu}\left(x,y\right)&=&\int\frac{d^{4}p}{(2\pi)^{4}}K_{(F)}^{\mu\alpha\nu\beta} \ \frac{p_{\alpha}p_{\beta}}{p^{4}}e^{-ip\cdot(x-y)} \ .
\end{eqnarray}

With the aid of Eq's (\ref{fields}), (\ref{propm}) and (\ref{proplv}) we can write 
\begin{eqnarray}
\label{fields2}
A^{\mu}\left(x\right)=A_{M}^{\mu}\left(x\right)+\Delta A_{LV}^{\mu}\left(x\right) \ .
\end{eqnarray}
where
\begin{eqnarray}
\label{defAs}
A_{M}^{\mu}\left(x\right)&=&\int d^{4}y D_{M}^{\mu\nu}\left(x,y\right)J_{\nu}(y)\cr\cr
\Delta A_{LV}^{\mu}\left(x\right)&=&\int d^{4}y D_{LV}^{\mu\nu}\left(x,y\right) J_{\nu}\left(y\right)
\end{eqnarray}

In a simplified setup, let us calculate the field configuration produced by a Dirac string (\ref{Dircurr2}). In this case, the first Eq. (\ref{defAs}) becomes
\begin{eqnarray}
\label{fieldM}
A_{M(D)}^{\mu}\left(x\right)=\frac{\Phi}{2\pi}\Bigl(0,\frac{-x^{2}}{(x^{1})^{2}+(x^{2})^{2}},\frac{x^{1}}{(x^{1})^{2}+(x^{2})^{2}},0\Bigr)\ . 
\end{eqnarray}
The potential (\ref{fieldM}) produces a vanishing electromagnetic field outside the $z$-axis.

For the Lorentz violation correction, we substitute Eq. (\ref{Dircurr2}) in the second Eq. (\ref{defAs}) and perform some simple manipulations,
\begin{eqnarray}
\label{fieldlv}
\Delta A_{LV(D)}^{\mu}\left(x\right)=\cr\cr
=\Phi \sum_{i,j=1}^{2}{\hat z}\cdot\Bigl[{\bf {\nabla}}_{{\bf x}_{\perp}}\times\Bigl(K_{(F)}^{\mu i1j}\ {\hat{x}}+K_{(F)}^{\mu i2j}\ {\hat{y}}\Bigr)\Bigr]\cr\cr
\times{\bf{\nabla}}_{{\bf x}_{\perp}}^{i}{\bf{\nabla}}_{{\bf x}_{\perp}}^{j}\lim_{m\rightarrow 0}\frac{1}{2m}\frac{\partial}{\partial m}\int\frac{d^{2}{\bf p}_{\perp}}{(2\pi)^{2}}\frac{e^{i{\bf p}_{\perp}\cdot{\bf x}_{\perp}}}{{\bf p}_{\perp}^2+m^{2}}\ ,\ \ \ \ \ \ \  
\end{eqnarray}
where ${\bf x}_{\perp}=\left(x^{1},x^{2},0\right)$.

The potential (\ref{fieldlv}) can be computed by following the same procedures employed in the previous section. The result is
\begin{eqnarray}
\label{fieldlv2}
\Delta A_{LV(D)}^{\mu}\left(x\right)=-\frac{\Phi}{2\pi{\bf x}_{\perp}^{2}}\left(x^{1}K_{(F)}^{\mu 121}-x^{2}K_{(F)}^{\mu 212}\right) \ .
\end{eqnarray}

From Eq. (\ref{fieldlv2}), we have a $0$-component for the field 
\begin{eqnarray}
\label{fieldlv3}
\Delta A_{LV(D)}^{0}\left(x\right)=-\frac{\Phi}{2\pi{\bf x}_{\perp}^{2}}\left(x^{1}K_{(F)}^{0121}-x^{2}K_{(F)}^{0 212}\right) \ , 
\end{eqnarray}
and a corresponding electric field outside the string, 
\begin{eqnarray}
\label{electric}
\Delta{\bf{E}}&=&-{\bf{\nabla}}_{{\bf x}_{\perp}}\left(\Delta A_{LV(D)}^{0}\right)\nonumber\\
&=&\frac{\Phi}{2\pi{\bf x}_{\perp}^{2}}\Biggl[-\frac{2}{|{\bf x}_{\perp}|}\left(x^{1}K_{(F)}^{0121}-x^{2}K_{(F)}^{0212}\right){\hat{x}}_{\perp}\nonumber\\
&
&+\left(K_{(F)}^{0121} \ {\hat{x}}-K_{(F)}^{0212} \ {\hat{y}}\right)\Biggr] \ ,
\end{eqnarray}
where ${\hat{x}_{\perp}}$ is a unit vector pointing in the direction of ${{\bf x}}_{\perp}$. 

From (\ref{fieldM}) and (\ref{fieldlv2}) the vector potential reads,
\begin{eqnarray}
\label{pv1}
{\bf{A}}_{(D)}&=&\frac{\Phi}{2\pi{\bf x}_{\perp}^{2}}\Bigl[\left(1-K_{(F)}^{1212}\right)\left(-x^{2}{\hat{x}}+x^{1}{\hat{y}}\right)\nonumber\\
&
&-\left(x^{1}K_{(F)}^{3121}-x^{2}K_{(F)}^{3212}\right){\hat{z}}\Bigr] \ .
\end{eqnarray}

From the vector potential (\ref{pv1}) we have an induced magnetic field outside the string, as follows 
\begin{eqnarray}
\label{magnetic}
\Delta{\bf{B}}&=&{\bf{\nabla}}_{{\bf x}_{\perp}}\times{\bf{A}}_{(D)}\nonumber\\
&=&-\frac{\Phi}{2\pi{\bf x}_{\perp}^{2}}\Biggl[-\Bigl(K_{(F)}^{3212} \ {\hat{x}}+K_{(F)}^{3121} \ {\hat{y}}\Bigr)\cr\cr
&\ &+\frac{2}{{\bf x}_{\perp}^{2}}\left(x^{1}K_{(F)}^{3121}-x^{2}K_{(F)}^{3212}\right)\left(-x^{2}{\hat{x}}+x^{1}{\hat{y}}\right)\Biggr] \ . \nonumber\\ 
\
\end{eqnarray}

The electric and magnetic fields, (\ref{electric}) and  (\ref{magnetic}), can induce physical phenomena outside the string.

Now, let us take a very simple and illustrative example for background tensor (which satisfies the conditions (\ref{simetriasQ}) and (\ref{dtrace})), as follows
\begin{eqnarray}
\label{parak71}
K_{(F)}^{0121}&=&K_{(F)}^{0212}=K_{(F)}^{3121}=K_{(F)}^{3212}=0 \ ,\nonumber\\ K_{(F)}^{1212}&=&k_{8} \ ,
\end{eqnarray}
where $k_{8}$ is a tiny constant. In this case, the $0$-component of the field  (\ref{fieldlv3}) vanishes and the vector potential (\ref{pv1}) becomes  
\begin{eqnarray}
\label{pv3}
{\bf{A}}_{(D)}&=&\frac{\Phi\left(1-k_{8}\right)}{2\pi{\bf x}_{\perp}^{2}}\left(-x^{2}{\hat{x}}+x^{1}{\hat{y}}\right) \cr\cr
&=&\frac{\Phi\left(1-k_{8}\right)}{2\pi\rho}{\hat{\phi}} \ ,
\end{eqnarray}
where we used cylindrical coordinates, with the radial coordinate $\rho=|{\bf x}_{\perp}|=\sqrt{(x^{1})^{2}+(x^{2})^{2}}$ and with $\hat{\phi}$ standing for the unitary vector for the azimuthal coordinate.

Notice that the potential (\ref{pv3}) does not produce any elctromagnetic field outside the string.

It is well known in the literature \cite{Grifthis,Sakurai} that the energy levels of a two dimensional quantum rigid rotor are modified when it circumvents an infinite solenoid. In this case we have a very simplified version of the so called Aharonov Bohm bound states \cite{Sakurai}. Taking a quantum rigid rotor composed by a non-relativistic particle with mass $M$ and electric charge $q$, restricted no move along a ring of radius $b$, adopting a coordinate system where the ring lies on the plane $z =0$, centered at the origin, considering a Dirac string placed along the $z$-axis, with internal magnetic flux $\Phi$, and using Eq. (\ref{pv3}) we can write the Hamiltonian for the charged particle in cylindrical coordinates
\begin{eqnarray}
\label{ahab1}
H=-\frac{1}{2Mb^{2}}\frac{d^{2}}{d\phi^{2}}+\frac{i q\Phi\left(1-k_{8}\right)}{2\pi Mb^{2}}\frac{d}{d\phi}\cr\cr
+\frac{q^{2}\Phi^{2}\left(1-k_{8}\right)^{2}}{8\pi^{2}Mb^{2}}
\end{eqnarray}
where it is implicit that we must discard the terms proportional to $k_{8}^{2}$.

The energy eigenfunctions of the hamiltonian (\ref{ahab1}) are given by
\begin{eqnarray}
\label{eigenfun}
\Psi\left(\phi\right)= Be^{in\phi}\ \ ,\ \ n=0,\pm1,\pm2,\cdots 
\end{eqnarray}
where $B$ is a normalization constant. Up to order $k_{8}$, the corresponding energy levels are
\begin{eqnarray}
\label{eneraha}
E_{n}=\frac{1}{2Mb^{2}}\left(n-\frac{q\Phi}{2\pi}\right)^{2}+\frac{q\Phi k_{8}}{2\pi Mb^{2}}\left(n-\frac{q\Phi}{2\pi}\right),
\end{eqnarray}

The first term on the right hand side of (\ref{eneraha}) is the well known Aharonov Bohm energy \cite{Grifthis} and the second term is a correction due to the Lorentz-symmetry breaking.


In terms of the SME coefficients, defined in Refs. \cite{CPT1,CPT2}, we can write  
\begin{eqnarray}
\label{k6}
k_{8}=\frac{1}{2}\left[(\tilde{\kappa}_{e+})^{33}-(\tilde{\kappa}_{e-})^{33}-{\tilde{\kappa}}_{{\mathrm{tr}}}\right] \ .
\end{eqnarray}

Therefore,
\begin{eqnarray}
\label{enerahaketr}
E_{n}&=&\frac{1}{2Mb^{2}}\left(n-\frac{q\Phi}{2\pi}\right)^{2}+\frac{q\Phi}{4\pi Mb^{2}}\left(n-\frac{q\Phi}{2\pi}\right)(\tilde{\kappa}_{e+})^{33}\nonumber\\
&
&-\frac{q\Phi}{4\pi Mb^{2}}\left(n-\frac{q\Phi}{2\pi}\right)\left[(\tilde{\kappa}_{e-})^{33}+{\tilde{\kappa}}_{{\mathrm{tr}}}\right]\ .
\end{eqnarray}

The energy (\ref{enerahaketr}) is due to the parity-even components of the background tensor. The second term on the right hand side is a birefringent contribution and the third one is nonbirefringent.


\section{Conclusions and perspectives}
\label{conclusoes}

In this paper we have investigated the interactions between external sources for the gauge field in the CPT-even sector of the SME. We have focused on physical phenomena which have no counterpart in Maxwell electrodynamics. We have obtained our results in $3+1$ dimensions and we treated the background tensor $K_{(F)\alpha\beta\sigma\tau}$ perturbatively up to first order. 

Specifically, we have showed that it emerges a spontaneous torque on a classical electromagnetic dipole and an interaction between a steady straight line current and a point-like charge. We have also investigated some phenomena due to the presence of a Dirac string. We have showed that the string can interact with a point charge as well as with a straight steady line current in the Lorentz symmetry breaking scenario. 

We have showed that our results are in agreement with the corresponding ones obtained in reference \cite{Fontes}, up to lowest order in the background vector, for the very restrictive situation where the background vector can be written in terms of just one single background vector (\ref{param}), the only one considered in Ref. \cite{Fontes}.

We have studied the so called Aharonov-Bohm bound states, for a $2$-dimensional quantum rigid rotator, in the Lorentz symmetry breaking scenario. We have obtained the energy levels for a specific example for the background tensor. 

The obtained results concerning deviations in the Coulombian behavior of the interaction energy between electric charges were used to estimate, euristically, upper bounds to the Lorentz-symmetry breaking parameters involved. The obtained estimates are far beyond the ones calculated with optical experimental data.

Some numerical
estimates have been made in order to investigate if some of the obtained effects were relevant for condensed
matter systems.

As a final remark, we point out that in this paper all the field sources considered are spinless. An interesting extension of this work would be the investigation of spin effects in the interactions between field sources.

\begin{acknowledgments}
L.H.C. Borges thanks to S\~ao Paulo Research Foundation (FAPESP) under the grant 2016/11137-5 for financial support. F.A. Barone thanks to CNPq (Brazilian agency) under the grants 311514/2015-4 and 313978/2018-2 for financial support.
\end{acknowledgments}



\end{document}